\newif\iflncs

\newcommand{\bref}[2]{\iflncs{#2} (see \cite{adams2023costs})\else\ref{#1}\fi}

\iflncs
\documentclass[runningheads,a4paper]{llncs} 

\else
\documentclass[letterpaper,11pt]{article}
\fi

\usepackage[utf8]{inputenc}
\usepackage[utf8]{inputenc}
\usepackage[ruled,vlined,linesnumbered]{algorithm2e}
\usepackage{xcolor}
\usepackage{amsthm}
\usepackage{braket}
\usepackage{amssymb}
\usepackage{amsmath}
\usepackage{xcolor}
\usepackage{soul}
\usepackage{mdframed}
\usepackage{enumitem}
\usepackage{mathtools}
\usepackage{hyperref}
\usepackage{boxedminipage}
\usepackage{graphicx}
\usepackage{subcaption}
\usepackage{booktabs}
\usepackage{multirow}

\usepackage{tabularray}
\usepackage{threeparttable}
\usepackage{float}
\usepackage{graphicx}
\usepackage{adjustbox}
\usepackage[labelfont=bf]{caption}
\usepackage{makecell}
\usepackage{flafter}

\iflncs
\setitemize{itemsep=0.25em}
\usepackage[pass]{geometry}
\else
\usepackage{geometry}
\fi
\usepackage{tabularx}
\usepackage{multirow}
\usepackage{rotating}

\iflncs
\title{Don't Let MEV Slip:\\The Costs of Swapping on the Uniswap Protocol}
    \titlerunning{Don't Let MEV Slip: The Costs of Swapping on the Uniswap Protocol}
    \author{Austin Adams\inst{1}\thanks{This work was made possible by mempool data made available to the authors by Blocknative (\url{https://www.blocknative.com}) and BloXroute (\url{https://bloxroute.com}). We are tremendously grateful for their generosity.}, Benjamin Y Chan\inst{2}\orcidID{0000-0003-1406-9845}\thanks{B. Chan---Work done while a Research Fellow at Uniswap Labs}, Sarit Markovich\inst{3}, \and
    Xin Wan\inst{1} }
    \institute{Uniswap Labs, New York, USA\\\email{austin@uniswap.org}, \email{xin@uniswap.org}
    \and
        Cornell Tech, New York, USA\\
    \email{byc@cs.cornell.edu}
    \and
    Northwestern University, Evanston, USA\\
    \email{s-markovich@kellogg.northwestern.edu}
    }
\else
\title{Don't Let MEV Slip:\\The Costs of Swapping on the Uniswap Protocol\thanks{This work was made possible by mempool data made available to the authors by Blocknative (\url{https://www.blocknative.com}) and BloXroute (\url{https://bloxroute.com}). We are tremendously grateful for their generosity.}}
\author{
Austin Adams\\Uniswap Labs\\\texttt{austin@uniswap.org} \and
Benjamin Y Chan\thanks{Work done while a Research Fellow at Uniswap Labs.}\\Cornell University\\\texttt{byc@cs.cornell.edu} \and Sarit Markovich\\Northwestern University\\\texttt{s-markovich@kellogg.northwestern.edu} \and Xin Wan\\Uniswap Labs\\\texttt{xin@uniswap.org}
}
\fi

\newcommand{\xin}[1]{{\color{blue} \textsf{Xin}: #1}}

\newcommand{\td}{\ensuremath{{\sf td}}}

\newcommand{\ignore}[1]{}

\newcommand{\E}{\mathbb{E}}

\usepackage{pifont}

\iflncs
    
    \spnewtheorem{claim}{Claim}{\itshape}{\rmfamily} 
    \spnewtheorem{subclaim}{Subclaim}{\itshape}{\rmfamily}

\else
    \usepackage{geometry}

\newtheorem{definition}{Definition}[section]

\newtheorem{protocol}{Protocol}[section]

\fi

\begin{document}

\maketitle

\begin{abstract}
We present the first in-depth empirical characterization of the costs of trading on a decentralized exchange (DEX). Using quoted prices from the Uniswap Labs interface for two pools --- USDC-ETH (5bps) and PEPE-ETH (30bps) --- we evaluate the efficiency of trading on DEXs. Our main tool is slippage --- the difference between the realized execution price of a trade, and its quoted price --- which we breakdown into its benign and adversarial components. We also present
an alternative way to quantify and identify slippage due to adversarial reordering of transactions, which we call \emph{reordering slippage}, that does not require quoted prices or mempool data to calculate. We find that the composition of transaction costs varies tremendously with the trade's characteristics. Specifically, while for small swaps, gas costs dominate costs, for large swaps price-impact and slippage account for the majority of it. Moreover, when trading PEPE, a popular ‘memecoin’, the probability of adversarial slippage is about 80\% higher than when trading a mature asset like USDC.

Overall, our results provide preliminary evidence that DEXs offer a compelling trust-less alternative to centralized exchanges for trading digital assets.
\end{abstract}

\iflncs
\else
\pagebreak
\fi
\section{Introduction}

Since its inception in 2018, the 
Uniswap Protocol \cite{adams2021uniswap,adams2023uniswap}---the largest \emph{decentralized exchange} by volume today---has handled nearly \$1.65 trillion (USD) in total notional transaction volume, or \$1.3 billion per day in 2023 alone. While decentralized exchanges (hereafter, DEXs) still see significantly less volume than traditional exchanges such as the NYSE and the CME, in part because trading is limited to assets that live on a blockchain, they point to a larger promise. It is a compelling promise: by using cryptography (such as digital signatures \cite{diffie1976new} or collision-resistant hash functions \cite{rabin1979digitalized,damgaard1987collision}, which power blockchains \cite{nakamoto2008bitcoin,buterin2014next}), we may design exchange protocols that allow us to transact without the need to trust a third party to offer a `fair' price or to settle trades. Moreover, they are `permissionless', allowing anyone to participate. In this way, decentralized exchanges promise to lower barriers to participation, boosting liquidity, whilst making trading fairer, auditable, and more efficient.

Whereas traditional exchanges typically route orders to a centralized matching engine---the opaqueness of which enables fraud \cite{mcmillanWired2014,gandal2018price,nytftxfraud2022}---DEXs take orders from a public mempool 
of pending transactions, before filling them on a blockchain using either a liquidity-pool/AMM \cite{buterin2016let,lu2017building} or a decentralized order book (e.g., EtherDelta). Still, as \cite{daian2020flash} first explored, moving things onchain does not necessarily guarantee fair and efficient trading. For example, block proposers have a monopoly on transaction ordering, due to requirements of an underlying consensus mechanism such as that of Ethereum \cite{buterin2020combining}.
This combined with the transparent nature of orders allows adversaries to carry out \emph{frontrunning} attacks, where they reorder transactions and insert their own orders (with the help of block proposers) to extract one form of `miner extractable value' (MEV) \cite{daian2020flash} from other users. Moreover, gas costs on Ethereum remain high, around approximately $\sim\$5$ to $\$25$ per trade on the Uniswap v3 in 2023. A user of a DEX will have to pay these costs (in addition to LP fees and price impact costs) on top of the market price whenever they transact. Quantifying these costs is important for evaluating existing exchanges and for designing new protocols.

To the best of our knowledge, few works (if any) have studied the overall transaction costs experienced by users of DEXs. 
The community has taken some initial steps \cite{0xslippage2022,blocknative2023} towards evaluating overall costs, but there remains a need for an in-depth analysis. \cite{barbon2023quality} estimate transaction costs of fixed-size trades over time by simulating a pool using on-chain liquidity and gas price data, but do not look at real trades or account for MEV or slippage.
Some past work quantified the magnitude of sandwich attacks, arbitrages, and liquidations on DEXs such as Uniswap v2 and v3 \cite{zhou2021high,qin2022quantifying,torres2021frontrunner}. However, they rely on heuristics to identify attacks, or on assumptions about the structure of MEV extraction \cite{modelingREV2022,mevexplore2022}, and may not capture other hidden sources of transaction costs. It is also essential to understand how the cost of MEV compares with other transaction costs, such as gas costs, LP fees, and price impact.

\paragraph{Summary of our contributions.}
In this paper, we study the overall efficiency of transacting on Uniswap v3, on Ethereum mainnet:
\begin{itemize}
    \item \emph{Framework.} We present a framework for evaluating the efficiency of DEXs. Our main tool is slippage --- the difference between the realized execution price of a trade, and its quoted price --- which provides a direct measure of the additional cost users pay on top of the quoted price.
    We also present an alternative way to quantify and identify slippage due to adversarial re-ordering of transactions, which we call \emph{reordering slippage}, that requires only on-chain data to compute and does not depend on heuristics. (While in this work, we have the luxury of access to mempool data and quoted prices, such data is usually proprietary and difficult to obtain.)
    
    \item \emph{Breakdown of transaction costs.} We analyze a dataset of 534,198 trades made through the Uniswap Labs interface for two pools that we believe are representative of trading on the Uniswap v3 --- the USDC-ETH (5bps) and PEPE-ETH (30bps) pools --- between January and mid-August, 2023. We present a detailed breakdown of transactions costs, the first such baseline that we know of.
    We find that while for small trades, gas fees dominate the transaction cost, for large trades, on average, slippage is by far the dominant cost. The average transaction cost per dollar transacted on USDC-ETH (PEPE-ETH) is about 22 bps (140 bps).
    Transaction costs vary widely depending on the pool and on the size of the trade.

    We show that, controlling for a variety of market variables, the effect of order size on slippage is significant, and that most of the effect of order size is driven by adversarial behavior. In addition, higher gas prices and market momentum worsen slippage, potentially due to collisions (i.e., if many users are simultaneously trading in the same direction).

    \item \emph{Other effects.} Slippage also strongly depends on the characteristics of the pool the user is trading against. We find that when trading PEPE, a popular `memecoin', the probability of adversarial slippage is historically about 80\% larger than when trading USDC, a mature asset. Finally, as an initial foray into evaluating the integrity of \texttt{mev-boost} participants, we find that private RPC services effectively eliminate adversarial slippage in our dataset (albeit this may change in the future). 
\end{itemize}
Our work establishes the first baseline for DEX efficiency.
Moreover, our techniques may provide a way to audit trusted entities (such as builders, relays, and RPC providers) in the \texttt{mev-boost} ecosystem to ensure that they are behaving honestly, a task previously thought to be difficult due to the opaque and centralized nature of block building. 

Looking forward, our results show early evidence that DEXs provide a sound trading experience, and offer a compelling trust-less alternative to centralized exchanges for trading digital assets. (The slippage we see is arguably \emph{lower} than what one would expect.) 
Unsurprisingly, our work also shows that there remains ample room for improvements, and for applying new techniques in cryptography and game theory to realize the vision of decentralized finance.

\iflncs
\section{Background}
The primary exchange that we analyze, Uniswap v3 \cite{adams2021uniswap,adams2023uniswap}, is an example of an Automated Market Maker (AMM), first explored in \cite{chen2012utility,buterin2016let}. We assume familiarity with basic financial terminology, Uniswap and AMMs, as well as with \texttt{mev-boost} \cite{mevboost2023} and the MEV ecosystem. For completeness, we describe AMMs and the MEV ecosystem, as well as terminology for working with Uniswap swaps, in Section 2 of the full paper \cite{adams2023costs}.

\else

\section{Background}
\label{sec:background}

\subsection{Decentralized Exchanges}

The primary exchange that we analyze, Uniswap v3 \cite{adams2021uniswap,adams2023uniswap}, is an example of an Automated Market Maker (AMM) \cite{buterin2016let}, where liquidity providers amass liquidity reserves $(x, y) \in \mathbb{R}^+ \times \mathbb{R}^+$ into a pool, where $x$ denotes the amount of the first asset and $y$ the amount of the second asset, and traders trade against the pool. In Uniswap v2, the pool reserves are constrained by the function
$$xy = k$$
for some constant $k$. Thus, a trader who wishes to purchase $y'$ units of the second asset, must deposit $x'$ units of the first asset s.t.
$$(x+x') \cdot (y - y') = k$$
in addition to paying settlement fees and additional fees set by the pool (which is 30 bps in Uniswap v2 or between 5 bps and 100 bps in Uniswap v3, which is then distributed pro-rata to liquidity providers, also known as the `LP fee'). Uniswap v3 additionally allows liquidity providers to specify a price range in which to deposit liquidity. For this paper, it suffices to note that the liquidity available to trade depends on the price; holding the available liquidity fixed, the protocol behaves like Uniswap v2. Automated market makers are typically implemented using a blockchain (in our case Ethereum), which manage the execution of the pool and keeps track of its state in a decentralized way. To trade with the pool, or deposit/withdraw liquidity, users send transactions to the underlying blockchain; their actions are realized only when the corresponding transaction is finalized on the blockchain. 

\subsection{Anatomy of a Swap}

A trade is a tuple comprising the \emph{input token}, or the ERC20 token that a user wishes to sell to the pool; the \emph{output token}, or the ERC20 token that the user wishes to receive in exchange from the pool; the \emph{fee tier}, identifying which pool to swap with; and one of either the \emph{input amount} or the \emph{output amount}, specifying the amount of the input token/output token that the user wishes to exchange (the amount of the other asset is then determined by the pool). 

A transaction for trading with a Uniswap pool is called a `swap', and comprises a trade  with two additional fields specified: one of
the \emph{minimum amount out} or the \emph{maximum amount in}, which is the worst case amount of the output/input asset that the user is willing to receive/spend; and a \emph{deadline}, specifying a deadline by which the swap must be completed, after which the swap is invalid. Even if a transaction is finalized on the blockchain, the underlying swap may fail due to a violation of the minimum amount out (resp. maximum amount in). For any given block $B$ in a blockchain, the trades contained in $B$ are defined to be the trades specified by the swaps in $B$ that succeed. 

When transacting on the Uniswap Labs interface, users are shown a quoted output amount (resp. input amount) for the input amount (resp. output amount) that they entered into the interface, in the form of a quoted average execution price. After seeing the quoted price, users can then decide whether to sign and broadcast the swap transaction.
The \emph{slippage tolerance} of a swap is defined as the ratio of the quoted amount out over the minimum amount out  (resp. the quoted amount in over the maximum amount in), minus 1, expressed in basis points (bps). The \emph{price impact} of a swap is defined as the ratio of the quoted price over the market mid price minus 1, expressed in bps.

\subsection{The MEV Ecosystem}

In order to make MEV extraction more transparent, Flashbots introduced the \texttt{mev-boost} library \cite{mevboost2023}, which is open source software that block proposers can run to connect them to a market of `block builders'. Block builders (which may eventually become enshrined in Ethereum through enshrined `proposer-builder separation' \cite{buterin2021proposer}) find the most profitable ways to sequence blocks of transactions, and bid for proposers to propose their blocks. The market is run by trusted relays, who send the block headers of the best bids to the block proposer; the proposer then replies with a signature of the winning block header, after which the relay sends the proposer the block (with the payment contained within). If the proposer later proposes a different block header than the one it signed initially, then it must have signed two different block headers and will be slashed. Additionally, block builders accept \emph{bundles} of transactions from `searchers' (users who want to avoid using the public mempool), and promise not to frontrun or unpack the transactions within bundles. Bundled transactions are also referred to as \emph{private} transactions.

Over 80\% of Ethereum validators today run a version of \texttt{mev-boost} \cite{mevexplore2022}. Note that \texttt{mev-boost} today requires many trust assumptions: proposers must trust relays to send them winning blocks, builders must trust relays to propagate their blocks/bids without publishing or frontrunning them, and searchers must trust builders not to publish or frontrun their private transactions. 

\fi

\section{Our Framework}
\label{sec:framework}

We present a framework for evaluating the execution quality and efficiency of DEXs. In particular, any such framework should generalize even as markets evolve (as is bound to happen, evidenced by the dominance of \texttt{mev-boost} since its introduction by Flashbots in 2020). Combined with the empirical results in Section~\ref{sec:empirical}, this gives a baseline for evaluating  DEXs as they become more sophisticated. 

In this paper, we focus on the following costs of trading --- slippage, settlement costs, exchange fees, liquidity costs --- and the latency of trades:
\begin{itemize}
    \item \emph{Slippage.}\footnote{In the literature, slippage is sometimes referred to as implementation shortfall \cite{perold1988implementation}.} 
    \label{def:slippage}
    Whereas many works have looked at dollar value lost specifically to sandwich attacks, backruns, etc. \cite{zhou2021high,torres2021frontrunner,qin2022quantifying,weintraub2022flash}, or dollar value earned by validators \cite{modelingREV2022,mevexplore2022}, to the best of our knowledge, we are the first to characterize the efficiency of a DEX by quantifying the overall slippage experienced by trades. 
    Following convention, we take positive slippage to denote a price improvement for the swapper:
\begin{definition}[Slippage] The slippage of swap $i$ (in bps) is
$$
\mathsf{slippage}_i := \left(\frac{\mathsf{realizedPrice}_i}{\mathsf{quotedPrice}_i} - 1\right) \cdot -10000 
$$
where $\mathsf{realizedPrice}_i$ is the average realized execution price of the swap (the amount of the input asset spent, over the amount of the output asset received), and $\mathsf{quotedPrice}_i$ is the decision price shown to the user.\footnote{Here, slippage is the difference between the realized output amount and the quoted output amount, expressed as a percentage of the realized output amount. Alternatively, we may express slippage as a percentage of the quoted output amount, with no major changes in interpretation.}
\end{definition}

\item \emph{Settlement costs, exchange fees.} When trading, users must pay for the operation of the exchange, as well as for settlement fees. On Ethereum, this takes the form of `gas' fees. Settlement and exchange fees are generally higher when volume is high and the underlying blockchain is congested; however, in Ethereum, gas costs scale sublinearly with the order size.

\item \emph{Liquidity fees and price impact.} For liquidity provision to be profitable, market makers take an explicit Liquidity Provider fee (LP fee) on AMMs
or construct a bid-ask spread on order books. 
The \emph{price impact} of a swap is defined as the ratio of the quoted price over the market mid price minus 1, expressed in bps, and directly measures market depth. We assume that the quoted price incorporates LP fees and the expected liquidity consumption of the swap, as in the case with the Uniswap Labs interface. 
\item \emph{Latency.} The latency of a trade is the time it takes the trade to be executed plus its settlement time; and should be minimized.
\end{itemize}
\ignore{
In addition to slippage (or execution quality) costs, a trader must pay exchange fees and transaction fees for each trade. When transacting with the Uniswap Protocol, users pay `gas fees' to the underlying blockchain (e.g. Ethereum) and a fee to the liquidity providers for each pool they transact with. Of course, market depth has the greatest influence on execution price; while decentralized exchanges like Uniswap v3 promise to broaden the liquidity supply by virtue of being permissionless (and by making passive strategies more competitive), an analysis of liquidity and market depth remains out of scope of this work.
}
We further break down slippage into its benign and adversarial components:

\iflncs

\paragraph{Adversarial slippage.} Adversarial slippage refers to slippage due to adversarial or reactive behavior. This includes frontruns, or more broadly, slippage due to MEV \cite{daian2020flash}. 
Usually, adversarial slippage is negative, but it may be positive, e.g., in the case of Just-in-Time liquidity provisioning (JIT) \cite{wan2022just}.

\paragraph{Collision slippage.} Collision slippage refers to slippage due to benign transactions sequenced between quote time and the final execution time. It may arise, for instance, if many traders are trying to trade the same assets at once on an exchange with non-zero latency.
\else
\begin{itemize}
\item \emph{Adversarial slippage.} Adversarial slippage refers to slippage due to adversarial or reactive behavior. This includes frontruns, or more broadly, slippage due to MEV \cite{daian2020flash}. 
Usually, adversarial slippage is negative, but it may be positive, e.g., in the case of Just-in-Time liquidity provisioning (JIT) \cite{wan2022just}.
\item \emph{Collision slippage.} Collision slippage refers to slippage due to benign transactions sequenced between quote time and the final execution time. It may arise, for instance, if many traders are trying to trade the same assets at once on an exchange with non-zero latency.
\end{itemize}
\fi

To measure adversarial and collision slippage, prior works relied on heuristics for classifying sandwich attacks or arbitrages \cite{qin2022quantifying,torres2021frontrunner}. In Section~\ref{sec:empirical}, we will use a different heuristic: we collect data from \texttt{mev-boost} to identify swaps that might be in the same MEV bundle and thus adversarial.

\iflncs
\paragraph{Other desiderata.} A DEX should be secure, decentralized (by principle), and fair. We will include discussion on these trade-offs when appropriate.
\else
\paragraph{Other desiderata.} 
The following desiderata should be kept in mind even though they do not directly impact the profit of market participants. We will include discussion on these trade-offs when appropriate:
\begin{itemize}
    \item \emph{Security.} Decentralized exchanges inherit their security from the underlying blockchain or consensus mechanism. As \cite{daian2020flash} first observed, DeFi protocols may even erode the security of underlying blockchains by providing new incentives to deviate from honest behavior at the consensus level \cite{gupta2023centralizing}.
    \item \emph{Decentralization.} Protocols such as \texttt{mev-boost} are centralized in that players must trust individual block builders and relays to be honest and not front-run private transactions. In the context of DeFi, decentralized solutions are preferable to centralized ones (by principle), and also promise additional accountability when compared to centralized exchanges.
    \item \emph{Fairness.} Exchanges should prioritize being \emph{permissionless} and avoid censoring any single user, preventing them from transacting or exchanging assets.
\end{itemize}
\fi

\subsection{Reordering Slippage}
Although a heuristical approach for identifying adversarial slippage makes sense for our empirical study, the notions of adversarial and collision slippage remain rather informal. A heuristical approach requires us to specify the exact attacks that we hope to measure (e.g. sandwich attacks or those captured by \texttt{mev-boost} payments), and will not capture unknown adversarial strategies; it also requires sophisticated mempool or \texttt{mev-boost} data that may be difficult to obtain.

As an alternative, we present a more formal notion which we call ``reordering slippage''. Reordering slippage is easy to measure without  mempool data, or even quoted prices, and also sidesteps the need for heuristics. 
Thus, it generalizes even as the market evolves with new adversarial strategies. 
Later, in Section~\ref{sec:empirical}, in addition to analyzing our heuristical notion of adverse slippage, we will also empirically analyze reordering slippage, and show that it indeed captures both sandwiches and arbitrages. 

\paragraph{Defining reordering slippage.} Reordering slippage compares the realized price of a swap to its hypothetical price in a world where the trades in its block are randomly ordered. Intuitively, if the realized price is far from this `randomly ordered' baseline, then the adversary must have explicitly `reordered' the trades in that block in order to extract value:
    \begin{definition}[Reordering Slippage] 
    \label{def:reorderingSlippage}
    Fix any block $B$, and denote by $S := \td_1, \ldots, \td_n$ the sequence of trades contained within $B$. For each trade $\td_i$, the reordering slippage (in bps) of its corresponding swap is defined as 
    $$\mathsf{reorderingSlippage}_{i} := \left( \frac{\mathsf{realizedPrice}_{i}}{\E_{\pi}[\mathsf{hypotheticalPrice}_{i}(\pi(S))]} - 1 \right) \cdot -10000
    $$
    where the expectation is taken over sampling a uniformly random permutation $\pi$ of the sequence of trades in $B$. Here, $\mathsf{realizedPrice}_i$ is the realized price for the $i$th trade, and $\mathsf{hypotheticalPrice}_{i}(\pi(S))$ refers to the hypothetical execution price of trade $\td_i$ if the trades in block $B$ are  ordered according to $\pi(S)$. \end{definition} 
    
    Note that when computing the hypothetical price, we consider a world where the actual realized trades (e.g., `sell 1 ETH') in $B$ are reordered, as opposed the swaps or transactions (e.g., `sell 1 ETH if we get at least 1800 USDC' back).\footnote{As an example, suppose a block contains three swaps, one of which fails due to a violation of its slippage tolerance setting, and two of which succeed. Suppose that the first successful trade $\td_1$---corresponding to some swap $s$---`buys 1 ETH'  and the second trade $\td_2$ `buys 2 ETH'. Then the reordering slippage of $s$ would be its realized price, divided by the average of its realized price and the hypothetical price of $\td_1$ in a world that first executes `buy 2 ETH' ($\td_2$) and then `buy 1 ETH' ($\td_1$). Note that even if $s$ would have failed (e.g., due to slippage tolerance) had it been ordered $2$nd, we can still compute a hypothetical price for executing $\td_1$ after $\td_2$.} Recall \iflncs\else from Section~\ref{sec:background}\fi that a trade is the realized interaction between a transaction and a pool. That is, only actual exchanges of assets factor into the calculation of reordering slippage, and, when reordered, they ``execute'' regardless of slippage tolerance settings or constraints set by a smart contract.\footnote{While the `hypothetical' world may not be realizable in practice, this definition allows us to quantify the effects of a broader set of adversarial strategies. In particular, those whose execution is conditional on the current state of the blockchain. To capture liquidity slippage, $S$ should include liquidity events in addition to trades.} 
        

\paragraph{Backruns and arbitrages.} Note that reordering slippage also captures backruns and arbitrages in the same block as the swap in question. While not directly adversarial, it seems reasonable to classify arbitrages as adversarial because arbitrages are costs born by liquidity providers \cite{milionis2022automated,milionis2023automated}, which are then passed on to users in the form of a larger LP fee.

\paragraph{Achieving zero reordering slippage.} Consensus-level order fairness \cite{kelkar2020order,heimbach2022sok} does not immediately guarantee zero reordering slippage. If transactions are ordered by observation time, backruns are possible and result in non-zero reordering slippage, pointing to market inefficiencies and LVR \cite{milionis2022automated}. 

A strawman approach to reduce reordering slippage is to randomize the order of transactions (or swaps) within each block (e.g., using an unpredictable randomness beacon). However, this also fails to achieve zero reordering slippage. The problem is that the execution of a transaction can depend on the current state of the blockchain. Thus, even if the order of transactions within a block is randomized and unpredictable, an adversarial transaction may simply refuse to carry through with a trade if executed in unfavorable conditions (via a smart contract, or simply by setting a tight slippage tolerance).\footnote{Indeed, this exact issue caused backrunning bots to spam the Ethereum network in the past \cite{livnev2020}.} As such, achieving zero reordering slippage remains an open question.

\paragraph{Concurrent work.} In concurrent work, \cite{angeris2023specter} present a notion that they call `Cost of MEV' which is very similar to our notion of reordering slippage (Definition~\ref{def:reorderingSlippage}). 
Their work is theoretical and 
shows loose bounds on how their worst-case notion scales with the total volume of trades, when instantiated with basic models for frontrunning and sandwiching. 
To use their notion still requires instantiating the metric with an appropriate model.
In contrast, we use reordering slippage to directly characterize the cost of trading on DEXs. Our choice of definition highlights the fact that even randomized transaction ordering does not eliminate adversarial slippage, since transactions can execute \emph{adaptively} based on their location in the block \cite{livnev2020}. Thus, the definition of reordering slippage seems well-equipped to capture MEV.
Moreover, we provide an empirical characterization of reordering slippage on Uniswap v3 in Section~\ref{sec:empirical}, quantifying realized execution costs (as opposed to the theoretical worst-case).

\section{Empirical Findings}
\label{sec:empirical}

\subsection{Data}

We obtain transaction hashes logged from the Uniswap Labs interface\footnote{\url{https://app.uniswap.org}}, covering all swaps made through the interface (mobile and web) that were published onchain, during the period of January 21, 2023 to August 14, 2023, and for two specific target pools on Ethereum: WETH-USDC 5bps pool (284,031 swaps) and WETH-PEPE 30bps pool (230,236 swaps).
For each swap, the dataset includes a `log index' which together uniquely identify the swap in question, the quoted price, and a timestamp for when the swap was relayed by the user. The quoted price is computed using onchain pool data at the time of the quote, and represents the expected execution price had the swap been executed at that time. In other words, the quoted price incorporates the estimated price impact of the swap and the LP fee. Due to caching, the quote may be one or two blocks stale.
Notably, our quoted prices comprise the actual prices shown to the user when they made the decision to swap on the interface.

We augment this dataset with additional onchain data to obtain the final average execution price, slippage tolerance setting, location in the blockchain, failure status, gas expenditure, and order size for each swap. We obtain onchain data for every swap in the target pools between January 21, 2023 to August 14, 2023, including but not limited to the interface swaps.\footnote{The data for every onchain swap will later be used to simulate Uniswap v3 pools to compute a slippage breakdown.} Onchain data is also used to compute the liquidity distribution of each target pool at the beginning of each block. We use the publicly available mevboost.pics dataset \cite{wahrstatter2023time} to obtain for each swap the builder that built the corresponding block.

We further augment our dataset with mempool data for every (onchain) swap, including the time that each swap was seen in the mempool, and whether a swap was seen for the first time in the public mempool, or if it was first seen as part of a finalized block onchain. The latter indicates that the swap was likely sent as part of a private \texttt{mev-boost} bundle of transactions. The mempool data comes from two different datasets: one assembled by bloXroute, comprising mempool data from June 22, 2023 to August 14, 2023, and one by Blocknative, comprising mempool data from January 21, 2023 to August 8, 2023. When mempool data overlaps, we take the earliest mempool observation time, and consider a swap public iff it was seen by both bloXroute and Blocknative in the public mempool. Most of the initial work was done using the bloXroute dataset, and subsequently extended to incorporate new data from Blocknative for the longer timeframe when it became available.

\paragraph{Pool choice.} We focus on two pools that we believe are representative of trading on DEXs on Ethereum mainnet --- the Uniswap v3 WETH-USDC\footnote{WETH is an ERC20 wrapped version of ETH backed 1-1. It is used in place of ETH due to most of DeFi requiring a token be an ERC20.} 5bps\footnote{pool address: 0x88e6A0c2dDD26FEEb64F039a2c41296FcB3f5640} and the Uniswap v3 WETH-PEPE 30 bps\footnote{pool address: 0x11950d141ecb863f01007add7d1a342041227b58} pools.
The WETH-USDC 5 bps pool is generally the largest pool by volume and liquidity in DeFi, trading mature assets that have a large market cap and good price discovery on centralized exchanges. PEPE is an archetypical high-volatility `memecoin' and representative of a less mature asset being traded on Uniswap v3; notably, the 30 bps pool has enough volume for a meaningful analysis. An exploration of a wider selection of pools is well-motivated and left to future work.

\subsection{Summary Statistics}
We study transaction costs from various angles: dollar amounts, fraction of trade sizes, as well as the breakdown into cost items laid out in Section~\ref{sec:framework}.

Our full sample includes 534,198 transactions, roughly \$6B dollar of volume. Combined, the sample accounts for approximately 20\% of all swaps done through the Uniswap Labs interface during the sample period, and roughly 12-15\% of the USD volume. The average WETH-USDC (WETH-PEPE) transaction size is \$18301.7 (\$2680.6), and the average total transaction cost (relative to a gas-free swap executed at the pool price at the end of the quote block) is \$40.7 (\$41.4), which is  22 bps (140 bps) of the mean order size. Note that this latter number represents the average transaction cost \emph{per dollar} transacted in each respective pool.
Swap sizes in our sample skew to the right, and the median swap size is only \$1,650 (\$407). For the median swap, gas cost is about \$7.3 (\$14.9), which is 44 bps (366 bps) of its order size.

\subsection{Cost Composition: High Level Patterns}
\label{sec:summary}

We compare the magnitude of the different transaction cost components: gas costs, slippage, LP fees, and the price impact of swaps. 

 
\begin{table}[]
    \centering
    \iflncs\fontsize{6}{7}\selectfont\else\fi
\begin{adjustbox}{max width = \dimexpr\paperwidth-3cm, center}
\begin{tabular}{lllllll}
\toprule
          &      & Total Cost &          Gas Cost &          Slippage &            LP Fee &        Price Impact \\
Token Pair & Size &                   &                   &                   &                     &            \\
\midrule
ETH$<>$USDC & All &  \$40.7 (22bps)&  \$10.1 (5bps) &  \$12.3 (7bps) &   \$9.3 (5bps) &     \$9.0 (5bps)  \\
         & Large & \$698.2 (24bps) &  \$16.9 (0.6bps) &  \$308.4 (11bps) &   \$145.4 (5bps) &     \$227.5 (8bps)  \\
          & Medium &  \$18.9 (14bps)  &  \$10.6 (8bps) &     \$1.0 (0.7bps) &     \$6.7 (5bps) &       \$0.6 (0.4bps) \\
          & Small & \$8.7 (250bps) &  \$8.5 (245bps) &     \$0.0 (0bps) &     \$0.2 (5bps) &       \$0.0 (0bps)  \\
ETH$<>$PEPE & All & \$41.4 (140bps)  & \$19.5 (66bps) &  \$3.0 (10bps) &  \$8.9 (30bps) &  \$10.1 (34bps) \\
        & Large & \$2847.9 (212bps) &   \$52.7 (4bps) &  \$190.8 (14bps) &  \$403.5 (30bps) &  \$2200.9 (164bps)  \\
          & Medium & \$67.9 (98bps) &  \$22.6 (33bps) &    \$8.3 (12bps) &   \$20.7 (30bps) &     \$16.3 (24bps)  \\
          & Small &  \$19.2 (526bps) & \$17.4 (478bps) &    \$0.5 (14bps) &    \$1.1 (30bps) &       \$0.2 (6bps)  \\
\bottomrule
\end{tabular}
\end{adjustbox}
    \vspace{1em}
    \caption{Mean Transaction Cost Composition by Pair \& Size; 
    Each dollar item is computed by total item cost divided by number of swaps in that bucket;
    each bps item is computed by total item cost divided by the total dollar volume for swaps in that bucket. Large Size group includes swaps larger than \$100,000; Medium Size group includes swaps between \$1,000 and \$100,000; Small Size Group includes swaps less than \$1,000. }
    \label{tab:unused}
\end{table}

Summing over all swaps, each of the four cost items accounts for between 20-35\% of the total transaction cost, with no item dominating costs, as shown in Figure~\bref{fig:pie_charts:fullsample}{1}. 
Yet, this breakdown varies with order size and across the different assets. Specifically, for the median sized transaction, gas cost completely dominates other cost, accounting for more than 90\% of total cost. Excluding gas costs, which are specific to the underlying blockchain, the overall cost of transacting in a decentralized market is on average 23bps (compared to 36bps including gas costs). Below we detail how the composition of total costs varies significantly with the size of the swap,  across different assets, and over time:

\paragraph{Order Size.} Transaction costs vary significantly with the size of the swap: the larger the swap, the lower the gas cost as a percentage of the swap size. In our WETH-USDC pool sample, gas cost is relatively tightly distributed around a median of \$7.3, with 75th and 25th percentile of \$5.0 and \$11.0, and a low correlation with swap size at $\sim$0.21. As a result, for small swaps with size below \$1000, over 98\% of total transaction cost is paid as gas cost; in contrast, for large swaps with size above \$100,000, gas cost accounts for 2.4\% of total transaction cost. That is, as swap size increases, other cost items start to dominate. In particular, for swaps larger than \$100,000, price-impact and slippage together account for  $\sim$77\% of overall transaction cost on a per dollar basis. This makes sense because price impact and slippage usually increase faster than swap size. Specifically, since marginal price impact is inversely related to liquidity depth, and Uniswap v3 pools tend to have thinner liquidity farther away from current pool price, then larger swaps should see increasing price-impact as a fraction of total order size. LP fees scale linearly with swap size, so their fraction of swap size remains fixed. 

\paragraph{USDC vs PEPE.} In our sample, the mean transaction cost for WETH-PEPE is about 140 bps, which is six times larger than the mean transaction cost for WETH-USDC swaps, at 22 bps. Several factors contribute to this large difference. PEPE is a relatively recent launched token, which means it has shallower liquidity, higher volatility, and smaller trade sizes. Specifically, the average and median transaction sizes for WETH-USDC pair is \$18,302 and \$1,650, respectively, while those of WETH-PEPE are only \$2,681 and \$407. Larger swap sizes mean that the relatively fixed gas cost will have a larger volume base to be amortized over, but at the same time adversaries like sandwich attackers could also have a strong motivation to attack. This is consistent with what we see in the data: for WETH-USDC swaps, summed slippage costs account for 30.1\% of total transaction cost\footnote{This is mainly driven by a few large swaps and is not a realistic average cost per swap.}, yet accounts for only 8.1\% for WETH-PEPE swaps. In contrast, gas costs account for 24.5\% of WETH-USDC cost, but 47.1\% of WETH-PEPE cost. Shallower liquidity also translates into higher price-impact and higher slippage. In our sample, the amount of liquidity within a 5\% price range ($\mathsf{liquidity}_i$) is on average 70-80 times deeper for the WETH-USDC pool. Finally, WETH-PEPE has higher price volatility, on average 10x times higher than WETH-USDC. As we will demonstrate below in our regression analysis, higher volatility, in general, leads to higher slippage cost. 

\paragraph{Time effects.} Transaction costs vary significantly based on the time period. We observe a significant reduction in total transaction cost as a fraction of trade size in the initial months of trading for WETH-PEPE. MEV related costs such as slippage and price-impact also spike during highly volatile periods. For the month of March, during the Silicon Valley Bank incident, the mean slippage and price-impact total 22.5 bps for the WETH-USDC pair and over 71\% of total transaction cost, more than five times the lowest point in the month of April, which saw a mean of 3.8 bps, accounting for only 25\% of total transaction cost.

\paragraph{Latency and fill rate.} We also present baseline data for latency and fill-rate of transactions. Latency and fill-rate measure how quickly and reliably can traders get into their desired positions. Our dataset indicates that roughly 90\% of transactions wait less than 12 seconds before their signed transactions are confirmed in a block. As we move out farther in the distribution, waiting time gets much longer---the 99.5th percentile waiting time is more than 20 blocks. Of the USDC-WETH swaps in our dataset, fewer than 0.5\% fail onchain. In PEPE-WETH, interface swaps saw an onchain fail rate of nearly 10\% at launch in mid-April, dropping to 5\% by the end of April, and approximately 3\% in August.

\paragraph{Comparison with Traditional Markets.} Many works have assessed transaction costs (slippage, commission, broker fees, bid ask spreads, price impact) on traditional markets.
Depending on the region and asset class, transaction costs can vary widely \cite{angel2015equity,chiyachantana2004international,domowitz2001liquidity}.
For example, \cite{angel2015equity} finds that transaction costs for equities range from 100 bps for emerging markets to 40 bps for US large cap companies. \cite{chiyachantana2004international} finds that institutional investors pay in the range of 40-70 bps in total transaction costs. The average transaction costs that we observe in the PEPE-WETH and USDC-WETH pools are remarkably competitive in comparison. 

\subsection{Factors Affecting Slippage}

We now shift our focus to understanding slippage on Uniswap v3, for two main reasons. First, of the four cost items above, gas used by a swap and LP fees are relatively static and do not change much with the swap's characteristics. 
In contrast, price-impact and slippage directly affect execution price on a per-trade basis. While the factors required to minimize price-impact are well understood, this is not the case for slippage. Second, slippage attracts the most adversarial attention. Thus, for any given swap, slippage arguably matters the most to execution quality.

We start by analyzing how different transaction and market factors affect slippage. To this end, we run a set of linear regressions on a broad set of market variables, for various measures of slippage, described by the equation
\begin{align}
\begin{split}
    \label{eq:mainregression}
        y_i = \beta_0 &+ \beta_1 \cdot \mathsf{orderSize}_i + \beta_2 \cdot \mathsf{gasPrice}_i + \beta_3 \cdot \mathsf{logLatency}_i\\ &+ \beta_4 \cdot \mathsf{slippageTolerance}_i + \beta_5 \cdot \mathsf{lastHourReturn}_i + \beta_6 \cdot \mathsf{liquidity}_i\\ &+ \beta_7 \cdot \mathsf{volatility}_i + \mathsf{weekFE} + e_i,
\end{split}
    \end{align}
where for each swap $i$, $y_i$ corresponds to one of the measures of slippage described above. In order to further control for market conditions, we control for weekly fixed effects. The results for the WETH-USDC 5 bps pool is presented in  Table~\ref{tab:overallregressions}, along with definitions for the above variables.
\iflncs 
\begin{table}
    \fontsize{6}{7}\selectfont
    \renewcommand\theadfont{}
\caption{\textbf{Factors Affecting Slippage (USDC-WETH 5 bps)}}
\vspace{1em}
    \label{tab:overallregressions}
{
\def\sym#1{\ifmmode^{#1}\else\(^{#1}\)\fi}
\begin{tabular}{@{\extracolsep{2pt}}l*{6}{c}@{}}
\hline\hline

 & \thead{\textsf{Slippage}\\(1)} & \thead{\textsf{AdversarialSlippage}\\(2)} & \thead{\textsf{CollisionSlippage}\\(3)} & \thead{\textsf{ReorderingSlippage}\\(4)} & \thead{\iflncs\textsf{TopOfBlockSg}\else\textsf{TopOfBlockSlippage}\fi\\(5)} & \thead{\textsf{LiquiditySlippage}\\(6)} \\
\hline
\textsf{orderSize} & -14.0176\sym{***} & -12.3729\sym{***} & -2.2378\sym{***} & -18.0085\sym{***} & -0.5170\sym{***} & 0.6208\sym{***} \\
 & (0.1670) & (0.0793) & (0.1511) & (0.0940) & (0.1306) & (0.0169) \\
\textsf{gasPrice} & -0.0043\sym{***} & -0.0004\sym{*} & -0.0038\sym{***} & -0.0011\sym{***} & -0.0028\sym{***} & -0.0001\sym{***} \\
 & (0.0004) & (0.0002) & (0.0003) & (0.0002) & (0.0003) & (0.0000) \\
\textsf{logLatency} & -0.0251\sym{+} & 0.0042 & -0.0251\sym{*} & 0.0091 & -0.0399\sym{***} & -0.0038\sym{**} \\
 & (0.0141) & (0.0067) & (0.0128) & (0.0079) & (0.0110) & (0.0014) \\
\textsf{slippageTolerance} & 0.0000\sym{+} & -0.0001\sym{***} & 0.0001\sym{***} & -0.0001\sym{***} & 0.0001\sym{***} & 0.0000 \\
 & (0.0000) & (0.0000) & (0.0000) & (0.0000) & (0.0000) & (0.0000) \\
\textsf{lastHourReturn} & -0.0115\sym{***} & 0.0008\sym{***} & -0.0123\sym{***} & -0.0004\sym{***} & -0.0087\sym{***} & 0.0000\sym{*} \\
 & (0.0002) & (0.0001) & (0.0002) & (0.0001) & (0.0002) & (0.0000) \\
\textsf{liquidity} & 0.0244\sym{***} & 0.0110\sym{***} & 0.0144\sym{***} & 0.0102\sym{***} & 0.0179\sym{***} & -0.0009\sym{**} \\
 & (0.0034) & (0.0016) & (0.0030) & (0.0019) & (0.0026) & (0.0003) \\
\textsf{volatility} & -0.0051\sym{***} & -0.0004 & -0.0046\sym{***} & -0.0022\sym{***} & -0.0022\sym{***} & -0.0001 \\
 & (0.0007) & (0.0003) & (0.0007) & (0.0004) & (0.0006) & (0.0001) \\
Intercept & -0.2951 & -0.0639 & -0.3057 & 0.0412 & -0.3399\sym{*} & 0.0726\sym{***} \\
 & (0.2164) & (0.1027) & (0.1958) & (0.1217) & (0.1692) & (0.0217) \\

\hline
Obs & 258925 & 258906 & 258906 & 258925 & 258925 & 255392 \\
Adj. R\sym{2} & 0.0471 & 0.0898 & 0.0235 & 0.1306 & 0.0152 & 0.0057 \\
F-stat & 356.5146 & 710.5045 & 174.4438 & 1081.6725 & 112.1161 & 41.6709 \\
\hline\hline
\end{tabular}
}

Every regression includes weekly fixed effects. ***, **, *, and $^+$ denote statistical significance at the 0.1\%, 1\%, 5\%, and 10\% levels.
The number of observations is less than reported in our summary statistics, due to some rows with missing latency or slippage tolerance information. All slippage notions are in units of bps.
\vspace{1em}

\emph{Variables Definition.}
The $\mathsf{Slippage}_i$ of swap $i$ is computed according to Definition~\ref{def:slippage}. Note that slippage for failed transactions (i.e. due to a violation of their slippage tolerance setting) is not well defined. Dropping these swaps from the analysis would bias the overall slippage. Instead, we simulate the execution price for each failed swap, and compute its hypothetical slippage in a world where the trade was executed. 
$\mathsf{AdversarialSlippage}_i$ and $\mathsf{CollisionSlippage}_i$ are calculated using a heuristic (see below). $\mathsf{ReorderingSlippage}_i$ is computed according to Definition~\ref{def:reorderingSlippage}. We calculate two additional slippage measures: $\mathsf{TopOfBlockSlippage}_i$ is the slippage calculated were the trade executed at the top of the block, comparing the top of block execution price to the quoted price, and $\mathsf{LiquiditySlippage}_i$ is the slippage attributable to transactions that deposit or withdraw liquidity in the same block as swap $i$. $\mathsf{orderSize}_i$ is the size of the swap, in units of millions of USDC. $\mathsf{gasPrice}_i$ is expressed in units of 1e-6 USDC, using the ETH price at the time of quote. $\mathsf{slippageTolerance}_i$ is the slippage tolerance setting of the swap, in non-negative basis points. The Uniswap Labs interface adds the slippage tolerance on top of the quoted price-impact of the swap to compute a minimum output (or input) amount. $\mathsf{liquidity}_i$ denotes the dollar value of the liquidity that would need to be consumed to increase the current price of the pool by 500 bps in the same direction as swap $i$.
$\mathsf{lastHourlyReturn}_i$ is the change in the pool price (in bps) in the hour prior to the corresponding swap, where a positive change corresponds to a worse price for the swapper (compared to if their trade executed an hour ago). $\mathsf{Volatility}_i$ is the standard deviation of one-minute log returns over a six hour sample. Finally, $\mathsf{logLatency}_i$ is the logarithm of the time the swap 'sat' in the mempool, that is, the time between when the swap is signed/broadcast by the user, and the time swap $i$ is finalized onchain.
\emph{Identifying Adversarial Slippage.}  
$\mathsf{AdversarialSlippage}_i$ for swap $i$ is computed using the following heuristic: first, consider the two transactions directly preceding $i$ in the same block $B$ (if any). For any trade executed within those two transactions, if the parent transaction is additionally private (i.e., part of an \texttt{mev-boost} bundle), then that trade is labeled `adversarial'. A `collision execution price' is then computed as the hypothetical average execution price of swap $i$, when starting with the pool state at the top of block $B$, then executing only the non-adversarial trades preceding swap $i$ in $B$, and then executing $i$'s trade. The $\mathsf{AdversarialSlippage}_i$ is the realized price (removing the effects of liquidity transactions in $B$) over the `collision execution price' (minus 1, times $-10000$). The $\mathsf{CollisionSlippage}_i$ is the `collision execution price' over the quoted price (minus 1, times $-10000$). A more principled heuristic is to label swaps in the same \texttt{mev-boost} bundle as $i$ as adversarial, but requires comprehensive bundle data across all builders, which is hard to obtain. Using a limited sample of Flashbots bundle data from June to August 2023, we verified that adversarial slippage computed using our heuristic closely tracks adversarial slippage computed using bundle data, for swaps in blocks built by Flashbots during said time period. 
 
\end{table} 
\else
\begin{table}
    \centering
    \footnotesize
\begin{adjustbox}{minipage=17.5cm, pagecenter}
\vspace{-60pt}
\caption{\textbf{Factors Affecting Slippage (USDC-WETH 5 bps)}}
\vspace{1em}
    \label{tab:overallregressions}

    Every regression includes weekly fixed effects. ***, **, *, and $^+$ denote statistical significance at the 0.1\%, 1\%, 5\%, and 10\% levels.
    The number of observations is less than reported in our summary statistics, due to some rows with missing latency or slippage tolerance information. All slippage notions are in units of bps.
 
\paragraph{Variables Definition.}
    
\emph{Identifying Adversarial Slippage.}  
$\mathsf{AdversarialSlippage}_i$ for swap $i$ is computed using the following heuristic: first, consider the two transactions directly preceding $i$ in the same block $B$ (if any). For any trade executed within those two transactions, if the parent transaction is additionally private (i.e., part of an \texttt{mev-boost} bundle), then that trade is labeled `adversarial'. A `collision execution price' is then computed as the hypothetical average execution price of swap $i$, when starting with the pool state at the top of block $B$, then executing only the non-adversarial trades preceding swap $i$ in $B$, and then executing $i$'s trade. The $\mathsf{AdversarialSlippage}_i$ is the realized price (removing the effects of liquidity transactions in $B$) over the `collision execution price' (minus 1, times $-10000$). The $\mathsf{CollisionSlippage}_i$ is the `collision execution price' over the quoted price (minus 1, times $-10000$). A more principled heuristic is to label swaps in the same \texttt{mev-boost} bundle as $i$ as adversarial, but requires comprehensive bundle data across all builders, which is hard to obtain. Using a limited sample of Flashbots bundle data from June to August 2023, we verified that adversarial slippage computed using our heuristic closely tracks adversarial slippage computed using bundle data, for swaps in blocks built by Flashbots during said time period.

\end{adjustbox}
\end{table}
\fi

Recall that for every notion of slippage that we consider, slippage is negative (by convention) if the realized execution price is worse than the quoted execution price, and positive if the realized price is better.

As Table~\ref{tab:overallregressions} shows, consistent with the high-level pattern observed in Section~\ref{sec:summary}, we observe in Model (1) that larger swaps (in terms of USDC size) are associated with substantially worse slippage: for every extra $1$ million dollars in additional order size, the swap costs an additional $14$ bps on top of its quoted price. That is, larger swaps pay a large penalty on top of their price-impact. While caution is needed when interpreting these numbers, this suggests that adversaries find larger swaps --- controlling for market volatility --- more profitable to sandwich or otherwise exploit (in the form of MEV). To test this, in models (2) \& (3) we run the same regression on the adverse and benign components of slippage. We break down slippage into adversarial, collision, and liquidity components (the methodology is presented in Table~\ref{tab:overallregressions}):
$$\mathsf{Slippage}_i \approx \mathsf{AdversarialSlippage}_i  + \mathsf{CollisionSlippage}_i  + \mathsf{LiquiditySlippage}_i $$

As expected, most of the effect of order size on slippage is driven by adversarial slippage. Moreover, increased liquidity reduces adversarial slippage. We also evaluate reordering slippage in model (4), and find that it behaves similarly to our heuristical notion of adverse slippage. 
Interestingly, the effect of order size on reordering slippage is larger than the effect in our original slippage and adversarial slippage models. This may be due to back-running transactions that are captured in reordering slippage, but not in adverse slippage. Note that when the number of swaps in the same block is high enough to make computing the reordering slippage computationally infeasible, we approximate it by sampling a constant number of permutations (e.g. 16) and taking the average. 

One interesting question is whether collision slippage is due to benign swaps in the same block as the target swap, or swaps from previous blocks (in which case latency might matter more).
As shown in model (5), putting the transaction at the top of the block does not completely eliminate slippage. While the coefficient is significantly smaller than the coefficient in models (1)-(4), it might take 2-3 blocks between the quote time and the time the transaction becomes onchain, so the price at the top of the block may be different than the quoted price (either due to collision or adversarial transactions), thereby creating slippage. Still, it seems like cross-block slippage is smaller than within-block slippage.

Another important factor affecting slippage is gas price. As the results show, higher gas prices are also negatively associated with slippage. Here, however, the association is mostly driven by collision slippage (see models 2 \& 3). The intuition is simple: gas price is higher when the networks is more congested and there is more activity. Since, typically, during high activity periods, users trade in the same directions, this results in higher negative slippage. This same intuition can explain the negative coefficient in models (4) \& (5).

Slippage is also expected to be affected by the time transactions sit in the mempool -- the longer it takes for the transaction to become onchain, the higher the expected slippage; due to higher probability of collision and higher probability of the transaction being identified as a profitable MEV opportunity. Our results show that indeed slippage increases with the time the transaction spends in the mempool. Interestingly, most of the effect comes from collision rather than adversarial slippage. This suggests that searchers respond quickly to MEV opportunities. As before, the negative effect in the case where the transaction is executed at the top of the block is due to inter-block slippage.

The hourly returns variable captures, in essence, the effect of market momentum. It is positive if the market moves in the same direction as the swap, and negative otherwise. As expected, market momentum negatively affects slippage and is mostly driven by collision. All other control variables -- e.g., volatility and liquidity -- are in the expected direction. Finally, note that the effect of the slippage tolerance set by the user is economically insignificant. This is driven by the fact that most users do not change the default tolerance of 50bps. Indeed, both the 25th percentile and the median slippage tolerance levels are 50bps.

\subsection{Comparison with the WETH-PEPE pool}

\iflncs
\emph{Note:} Due to space constraints, some referenced Tables and Figures are found only in the full version of the paper \cite{adams2023costs}.
\fi

As mentioned above, the WETH-USDC pool is a mature and highly liquid pool. One would expect it to perform much more efficiently than younger, mostly speculative pools with high volatility. To this end, we choose the WETH-PEPE pool which is a highly active pool with $1/100$ the liquidity of USDC (median of $0.24M$ vs $22.5M$ dollars) and about $10$ times the volatility of USDC. Table~\bref{tab:pepeoverall}{3} presents the regression results.

In general, the direction of the effects of the different factors on slippage is similar to what we see in the WETH-USDC pool, with the notable exception that the effect of gas price on adversarial slippage is positive and significant and the effect of slippage tolerance is significant and economically meaningful. The positive and significant coefficient on gas price for adversarial slippage is likely due to the fact that an increase in the cost of a transaction makes some adversarial strategies unprofitable. If, on average, the availability of profitable MEV opportunities does not change during high gas price, then we should expect to see less adversarial activity during times when gas price is high. Since, unlike the WETH-USDC pool, network congestion is likely not closely associated with profitable MEV opportunities in the PEPE pool, we see that adversarial slippage for PEPE is positively correlated with gas price. As for slippage tolerance, the 25th percentile and median slippage tolerance values for the PEPE pool are 100 and 300, respectively. This suggests that users (or the Uniswap Labs interface) are actively choosing risk tolerance levels with the expectation that slippage would be quite high, likely, due to the high price volatility.

Given the large differences in overall activity, transaction size, etc. between the two pools, the coefficients in the regressions in Table~\ref{tab:overallregressions} and Table~\bref{tab:pepeoverall}{3} cannot be directly compared. In order to examine whether there is more adversarial activity in the PEPE pool relative to the USDC pool, we run a logit regression on adversarial slippage. Specifically, we run the same regression as in Equation~\ref{eq:mainregression} where now $y_i$ takes on the value of $1$ if swap $i$ experienced a negative adversarial slippage larger than $\$5$, and $0$ otherwise.\footnote{To avoid the case of swaps being mistakenly identified as adverse because of a small negative adversarial slippage (which may be spurious due to our heuristic), we classify transactions in the logit regression as being adversarial only if their adversarial slippage (in bps) multiplied by the order size is worse than negative $\$5$.} The results are presented in Table~\bref{tab:probitreg}{4}. (Note that liquidity in range is quite correlated with week number for PEPE, and much less so for USDC).

As the table shows, the likelihood of adversarial slippage for PEPE is about $80\%$ larger than for USDC. Furthermore, the increased likelihood is enhanced by the size of the transaction--i.e., for a certain increase in transaction size, the increase in the likelihood of adversarial slippage for PEPE is larger than the corresponding increase for USDC.

The results above suggest that slippage in mature markets that are highly active, liquid, and with low volatility is minor. More generally, the results suggests that, despite the decentralization and transparency that characterize DeFi market, mature markets behave efficiently and may be even considered to be close in their efficiency to traditional financial markets. In order to further examine the impact of the fundamental characteristics of DeFi market---decentralization and transparency---next, we break down our analysis to public and private transactions. Specifically, while decentralization is core to the functioning of the Uniswap Protocol, nowadays many transactions (including MEV transactions) are sent as private transactions. This allows us to better study the effect of transparency on market efficiency in mature markets as WETH-USDC as well as in younger markets like the WETH-PEPE pool.   

\section{The Impact of MEV Infrastructure on Slippage}

We examine two important components of the MEV ecosystem: the usage of private RPCs and trust in builders' neutrality.

\subsection{Private RPCs}

Sending a transaction to a private RPC (putting it in a \texttt{mev-boost} bundle) should prevent other searchers from sandwiching transactions, as the private transaction does not appear in the mempool before becoming onchain. Consequently, private transactions should have much smaller negative slippage, that is only driven by collision. Nevertheless, since these transactions typically are sent to specific RPCs, they might suffer from higher latency and consequently wider collision slippage.

Table~\bref{tab:privateregressions}{5} presents the  results for the same regression as in Equation~\ref{eq:mainregression} with the addition of a dummy variable  that takes on the value of $1$ if transaction $i$ is public, and $0$ otherwise. As expected, the coefficient on $\textsf{Public}$ is negative and significant for both USDC and PEPE. In fact, out of the 8294 (8629) private interface swaps in USDC (PEPE), only 3 (12) have negative adverse slippage that is larger than \$5. That is, private RPCs seem to completely eliminate adverse slippage. Of the 3 (12) swaps with adverse slippage worse than $\$5$, none are obviously sandwiched when manually checked on Etherscan, and the slippage appears accidental. Furthermore, the coefficient on collision slippage is not significant, meaning that the potential increase in latency has no effect on average collision slippage. We note that while this shows evidence that private RPCS are reliable in the present day, it does not guarantee that the trust assumptions that underpin private RPCs will continue to be valid in the future.\footnote{We further interact $\mathsf{Public}$ with our other explanatory and control variables. We find that the effect of order size on adversarial slippage is stronger for public transactions, yet gas price interacted with $\mathsf{Public}$ has no significant effect. For brevity, we do not present these results here.} 

\subsection{Builder Trust}

When participating in the \texttt{mev-boost} ecosystem, searchers and users of private RPCs must trust that builders do not frontrun or `unpack' the \texttt{mev-boost} bundles that are sent to builders. While in traditional markets, it is the regulator that audits the intermediaries, in DeFi this trust relies on incentives (or even on goodwill). It is, therefore, important to audit this trust assumption. Our reordering slippage provides a way for the public to monitor builders' behavior, without the need to acquire private data.

As Table~\bref{tab:builders}{6} shows, we do not find conclusive evidence that any of the top 5 builders (by private transaction count) are misbehaving, at least from a cursory examination.  This may suggest that the penalties associated with breaking users' trust are large enough to incentivize builders not to defect. Investigating the validity of trust assumptions required by the MEV ecosystem remains an important open question.

\ignore{Another metric of concern that may negatively affect trust is the degree of concentration in the block building market. This is worth monitoring because it highlights the centralization external to the core Ethereum protocol. \xin{Ben to check language here} Our dataset highlights the degree of existing centralization as well as the potential directional impact of future changes on the centralization tendency. Currently, the top five builders in aggregate produce blocks that account for roughly 85\% - 92\% of swaps by count and about 93\% of USD volume in our dataset.
}

\iflncs

\section{Related work}
\label{sec:relatedwork}

Quantifying MEV on Ethereum is an active area of research. Daian et al. \cite{bentov2017cost,daian2020flash} introduced the notion of MEV and were the first to demonstrate its impact on decentralized markets, in turn giving rise to projects such as \texttt{mev-boost} \cite{mevboost2023}. More recently, a series of works \cite{torres2021frontrunner,qin2022quantifying} quantify frontrunning attacks and MEV extraction using historical data. \cite{weintraub2022flash} further extend this to analyze the impact of Flashbots on overall MEV extraction. Many of their techniques rely on identifying specific strategies for extracting value from users, and then designing heuristics for identifying those strategies. More recently, \cite{qin2021attacking,babel2023clockwork,babel2023lanturn} explore more automated approaches for identifying value-extraction mechanisms.

Few works have analyzed slippage or overall transaction costs of trading on decentralized exchanges. 0x \cite{0xslippage2022} analyze the slippage of trades sent through the 0x Swap API, and show that setting an appropriate slippage tolerance is essential for bounding the cost of MEV experienced by any single swap. \cite{blocknative2023} show that transactions sent to private RPCs may have on average higher transaction costs than public transactions. \cite{barbon2023quality} estimate the transaction costs for various fixed-size trades over time by simulating pools using onchain liquidity data, and gas price data. Their analysis does not incorporate slippage, MEV, or onchain execution prices, relying entirely on simulated execution.

A number of works measure the dynamics of liquidity provisioning on decentralized exchanges, sometimes through the lenses of transaction costs. \cite{caparros2023blockchain} show that lower gas fees increase liquidity repositioning and concentration, reducing price impact for small trades. They use a notion of slippage that incorporates price impact, comparing realized prices to the market mid price. \cite{hasbrouck2022need} show that higher LP fees may reduce price impact. 
\cite{liao2022uniswapv3} show that the cost of price impact on AMMS may be lower than on centralized exchanges for highly liquid pools.

\emph{Transaction costs on traditional markets.} In contrast, a number of works have quantified overall trading costs on traditional markets (such as equities markets). We point to 
\cite{domowitz1999automation,domowitz2001liquidity,domowitz2001global,chiyachantana2004international,angel2015equity} as examples, but this list is by no means comprehensive.

\emph{Mitigating MEV.} Much work has focused on mitigating MEV extraction through protocol design. At the consensus layer, \cite{kelkar2020order} propose an order-fair consensus algorithm, where transactions are ordered in the time they arrive in the view of validators. Order-fair consensus has seen substantial follow-up work \cite{zhang2020byzantine,kursawe2021wendy,cachin2022quick,kelkar2022order}. Time-based order-fairness may not eliminate backruns or arbitrage and may incentivize latency wars. An alternative approach, as described by \cite{malkhi2022maximal}, is to force block proposers to `commit' to an ordering of a block, before they learn the contents of the transactions. Such an approach is reminiscent of those used by MPC protocols \cite{yao1986generate,micali1987play} and asynchronous byzantine agreement algorithms; the downside is that it requires more sophisticated cryptography (e.g. threshold cryptography, secret sharing) that is harder to adapt to a permissionless setting.

At the block builder level,  \cite{xavier2023credible} propose a verifiable block sequencing rule that block builders can follow and that observers can audit. Such a rule may mitigate MEV whilst being accountable to the general public.




\else

\pagebreak
\fi

\bibliographystyle{splncs04}
\bibliography{main}

\appendix

\iflncs
\fi



\iflncs
\else
\section{Tables and Regressions}
\label{sec:morecharts}

In this section, we present additional tables that are used in Section~\ref{sec:empirical}.

\begin{table}
\centering
\footnotesize
\begin{adjustbox}{minipage=17.5cm, pagecenter}
\caption{\textbf{Factors Affecting Slippage (PEPE-WETH 30 bps)}}
\vspace{1em}
    \label{tab:pepeoverall}
{
\def\sym#1{\ifmmode^{#1}\else\(^{#1}\)\fi}
\begin{tabular}{@{\extracolsep{2pt}}l*{6}{c}@{}}
\hline\hline

 & \thead{\textsf{Slippage}\\(1)} & \thead{\textsf{AdversarialSlippage}\\(2)} & \thead{\textsf{CollisionSlippage}\\(3)} & \thead{\textsf{ReorderingSlippage}\\(4)} & \thead{\textsf{TopOfBlockSlippage}\\(5)} & \thead{\textsf{LiquiditySlippage}\\(6)} \\
\hline
\textsf{orderSize} & -240.5533\sym{***} & -298.8950\sym{***} & 3.3267 & -407.8857\sym{***} & -2.2687 & 66.8170\sym{***} \\
 & (22.8863) & (10.4824) & (20.8326) & (12.6344) & (18.4876) & (3.2512) \\
\textsf{gasPrice} & -0.0427\sym{***} & 0.0038\sym{*} & -0.0479\sym{***} & -0.0060\sym{**} & -0.0333\sym{***} & 0.0014\sym{*} \\
 & (0.0040) & (0.0018) & (0.0036) & (0.0022) & (0.0032) & (0.0006) \\
\textsf{logLatency} & -0.7973\sym{***} & 0.0882 & -0.8462\sym{***} & 0.4903\sym{***} & -1.6148\sym{***} & -0.0589\sym{+} \\
 & (0.2324) & (0.1064) & (0.2116) & (0.1283) & (0.1877) & (0.0334) \\
\textsf{slippageTolerance} & -0.0117\sym{***} & -0.0053\sym{***} & -0.0067\sym{***} & -0.0058\sym{***} & -0.0058\sym{***} & 0.0002\sym{**} \\
 & (0.0004) & (0.0002) & (0.0004) & (0.0002) & (0.0004) & (0.0001) \\
\textsf{lastHourReturn} & -0.0142\sym{***} & -0.0011\sym{***} & -0.0137\sym{***} & -0.0015\sym{***} & -0.0103\sym{***} & 0.0007\sym{***} \\
 & (0.0004) & (0.0002) & (0.0004) & (0.0002) & (0.0004) & (0.0001) \\
\textsf{liquidity} & 21.8502\sym{***} & 3.8909\sym{***} & 20.3436\sym{***} & 6.9135\sym{***} & 13.8188\sym{***} & -2.1611\sym{***} \\
 & (2.4339) & (1.1150) & (2.2160) & (1.3436) & (1.9661) & (0.3442) \\
\textsf{volatility} & -0.0259\sym{***} & -0.0019\sym{***} & -0.0244\sym{***} & -0.0051\sym{***} & -0.0177\sym{***} & 0.0007\sym{***} \\
 & (0.0010) & (0.0005) & (0.0009) & (0.0006) & (0.0008) & (0.0001) \\
Intercept & -26.8079\sym{***} & -12.1405\sym{***} & -12.9395\sym{***} & -15.4330\sym{***} & -8.6243\sym{***} & -0.7573\sym{**} \\
 & (1.7050) & (0.7809) & (1.5520) & (0.9413) & (1.3773) & (0.2469) \\

\hline
Obs & 170417 & 170405 & 170405 & 170417 & 170417 & 162493 \\
Adj. R\sym{2} & 0.0309 & 0.0123 & 0.0257 & 0.0158 & 0.0179 & 0.0049 \\
F-stat & 237.0294 & 93.4615 & 196.3155 & 120.1161 & 136.0163 & 35.6470 \\
\hline\hline
\end{tabular}
}

    Every regression includes weekly fixed effects. ***, **, *, and $^+$ denote statistical significance at the 0.1\%, 1\%, 5\%, and 10\% levels. The methodology is the same as for the USDC-WETH analysis in Table~\ref{tab:overallregressions}. The number of observations is less than reported in our summary statistics, due to some rows with missing latency or slippage tolerance information, inherited from the quality of our interface data.
\end{adjustbox}
\end{table}

\begin{table}
\centering
\footnotesize
\begin{adjustbox}{minipage=14.1cm, pagecenter}
\vspace{-60pt}
\caption{\textbf{Comparing PEPE-WETH with USDC-WETH}}
\vspace{1em}
     \label{tab:probitreg}
{
\def\sym#1{\ifmmode^{#1}\else\(^{#1}\)\fi}
\begin{tabular}{@{\extracolsep{2pt}}l*{3}{c}@{}}
\hline\hline

 & \thead{($\textsf{AdverseSlippage}_i < -5$)\\ USDC only (1)} & \thead{($\textsf{AdverseSlippage}_i < -5$)\\ PEPE only (2)} & \thead{($\textsf{AdverseSlippage}_i < -5$)\\ combined (3)} \\
\hline
\textsf{orderSize} & 7.0657\sym{***} & 37.1310\sym{***} & 7.1853\sym{***} \\
 & (0.1335) & (0.9355) & (0.2973) \\
\textsf{gasPrice} & -0.0011\sym{+} & 0.0003 & 0.0031 \\
 & (0.0006) & (0.0003) & (0.0025) \\
\textsf{logLatency} & -0.0624 & -0.0451\sym{+} & 0.0354 \\
 & (0.0430) & (0.0246) & (0.1012) \\
\textsf{slippageTolerance} & -0.0006\sym{**} & -0.0001 & -0.0041\sym{**} \\
 & (0.0002) & (0.0000) & (0.0015) \\
\textsf{lastHourReturn} & -0.0006\sym{+} & 0.0003\sym{***} & 0.0017 \\
 & (0.0004) & (0.0000) & (0.0010) \\
\textsf{liquidity} & -0.1109\sym{***} & -4.2594\sym{***} & -0.1175\sym{***} \\
 & (0.0102) & (0.2938) & (0.0217) \\
\textsf{volatility} & 0.0087\sym{***} & 0.0003\sym{***} & 0.0121\sym{*} \\
 & (0.0017) & (0.0001) & (0.0055) \\
\textsf{isPepe} &  &  & 2.0327\sym{*} \\
 &  &  & (0.9200) \\
\textsf{isPepe}:\textsf{orderSize} &  &  & 29.9457\sym{***} \\
 &  &  & (0.9816) \\
\textsf{isPepe}:\textsf{gasPrice} &  &  & -0.0027 \\
 &  &  & (0.0025) \\
\textsf{isPepe}:\textsf{logLatency} &  &  & -0.0805 \\
 &  &  & (0.1042) \\
\textsf{isPepe}:\textsf{slippageTolerance} &  &  & 0.0040\sym{**} \\
 &  &  & (0.0015) \\
\textsf{isPepe}:\textsf{lastHourReturn} &  &  & -0.0014 \\
 &  &  & (0.0010) \\
\textsf{isPepe}:\textsf{liquidity} &  &  & -4.1419\sym{***} \\
 &  &  & (0.2946) \\
\textsf{isPepe}:\textsf{volatility} &  &  & -0.0117\sym{*} \\
 &  &  & (0.0055) \\
Intercept & -4.6234\sym{***} & -4.5203\sym{***} & -6.5530\sym{***} \\
 & (0.2830) & (0.1342) & (0.9101) \\

\hline
Obs & 258906 & 170405 & 276177 \\
Adj. R\sym{2} &  &  &  \\
F-stat &  &  &  \\
\hline\hline
\end{tabular}
}
     
     This table presents the coefficients of running a logistic regression to estimate the following model:
$$
y_i=\begin{cases}
			1 & \text{if $w_i > 0$}\\
            0 & \text{otherwise}
\end{cases}$$
\vspace{-1em}
\begin{align*}
\begin{split}
        w_i = \beta_0 &+ \beta_1 \cdot \mathsf{orderSize}_i + \beta_2 \cdot \mathsf{gasPrice}_i + \beta_3 \cdot \mathsf{logLatency}_i + \beta_4 \cdot \mathsf{slippageTolerance}_i\\ &+ \beta_5 \cdot \mathsf{lastHourReturn}_i + \beta_6 \cdot \mathsf{liquidity}_i + \beta_7 \cdot \mathsf{volatility}_i + \mathsf{biweekFE} + e_i,
\end{split}
\end{align*}
where $w_i$ is an unobserved latent variable. 
Here, $y_i$ is a dummy variable that is equal to
1 if $\textsf{AdversarialSlippage}_i < -5$ and is 0 otherwise. Model (1) shows the results for only swaps in the USDC-WETH 5 bps pool. Model (2) shows the results for the PEPE-WETH 30 bps pool. In model (3), we further interact every term with a dummy variable $\textsf{isPepe}_i$ denoting whether a swap came from the USDC or the PEPE dataset.
Models (1) and (2) include biweekly fixed effects (for every two weeks). Model (3) includes a fixed effect for each pool/biweek pair. The variables are defined as in Table~\ref{tab:overallregressions}. 
\end{adjustbox}
\end{table}

\begin{table}
\centering
\scriptsize
\begin{adjustbox}{minipage=17cm, pagecenter}
\vspace{-60pt}
\caption{\textbf{Evaluating Private RPCs}}
\vspace{1em}
        \label{tab:privateregressions}
{
\def\sym#1{\ifmmode^{#1}\else\(^{#1}\)\fi}
\begin{tabular}{@{\extracolsep{2pt}}l*{6}{c}@{}}
\hline\hline

 & \thead{\textsf{Slippage}\\ USDC only (1)} & \thead{\textsf{AdversarialSlippage}\\ USDC only (2)} & \thead{\textsf{CollisionSlippage}\\ USDC only (3)} & \thead{\textsf{Slippage}\\ PEPE only (4)} & \thead{\textsf{AdversarialSlippage}\\ PEPE only (5)} & \thead{\textsf{CollisionSlippage}\\ PEPE only (6)} \\
\hline
\textsf{Public} & -0.3179\sym{**} & -0.1892\sym{***} & -0.1472\sym{+} & -10.2028\sym{***} & -4.3599\sym{***} & -6.8655\sym{***} \\
 & (0.0969) & (0.0460) & (0.0876) & (1.4884) & (0.6817) & (1.3548) \\
\textsf{orderSize} & -14.0278\sym{***} & -12.3789\sym{***} & -2.2425\sym{***} & -254.0085\sym{***} & -304.6394\sym{***} & -5.7190 \\
 & (0.1671) & (0.0793) & (0.1511) & (22.9679) & (10.5195) & (20.9075) \\
\textsf{gasPrice} & -0.0043\sym{***} & -0.0004\sym{*} & -0.0038\sym{***} & -0.0432\sym{***} & 0.0036\sym{+} & -0.0482\sym{***} \\
 & (0.0004) & (0.0002) & (0.0003) & (0.0040) & (0.0018) & (0.0036) \\
\textsf{logLatency} & -0.0285\sym{*} & 0.0020 & -0.0267\sym{*} & -1.0264\sym{***} & -0.0098 & -1.0006\sym{***} \\
 & (0.0142) & (0.0067) & (0.0128) & (0.2348) & (0.1075) & (0.2137) \\
\textsf{slippageTolerance} & 0.0000 & -0.0001\sym{***} & 0.0001\sym{***} & -0.0120\sym{***} & -0.0054\sym{***} & -0.0069\sym{***} \\
 & (0.0000) & (0.0000) & (0.0000) & (0.0004) & (0.0002) & (0.0004) \\
\textsf{lastHourReturn} & -0.0115\sym{***} & 0.0008\sym{***} & -0.0123\sym{***} & -0.0142\sym{***} & -0.0011\sym{***} & -0.0137\sym{***} \\
 & (0.0002) & (0.0001) & (0.0002) & (0.0004) & (0.0002) & (0.0004) \\
\textsf{liquidity} & 0.0244\sym{***} & 0.0110\sym{***} & 0.0144\sym{***} & 21.9822\sym{***} & 3.9439\sym{***} & 20.4271\sym{***} \\
 & (0.0034) & (0.0016) & (0.0030) & (2.4342) & (1.1149) & (2.2159) \\
\textsf{volatility} & -0.0051\sym{***} & -0.0004 & -0.0046\sym{***} & -0.0259\sym{***} & -0.0019\sym{***} & -0.0244\sym{***} \\
 & (0.0007) & (0.0003) & (0.0007) & (0.0010) & (0.0005) & (0.0009) \\
Intercept & 0.0295 & 0.1294 & -0.1553 & -16.1428\sym{***} & -7.5826\sym{***} & -5.7623\sym{**} \\
 & (0.2380) & (0.1129) & (0.2153) & (2.3081) & (1.0571) & (2.1011) \\

\hline
Obs & 258906 & 258906 & 258906 & 170405 & 170405 & 170405 \\
Adj. R\sym{2} & 0.0471 & 0.0899 & 0.0236 & 0.0311 & 0.0126 & 0.0258 \\
F-stat & 347.1956 & 691.8020 & 169.8065 & 229.1725 & 91.2927 & 189.2329 \\
\hline\hline
\end{tabular}
}
\footnotesize

In this table, we present coefficients for regressions following Equation~\ref{eq:mainregression} but with an extra dummy variable $\mathsf{Public}_i$ which denotes whether swap $i$ is first seen in the public mempool (else we consider it a private swap). Models (1), (2), and (3) are run for the USDC-WETH dataset, and models (4), (5), and (6) are run for the PEPE-WETH dataset. Every regression includes weekly fixed effects. ***, **, *, and $^+$ denote statistical significance at the 0.1\%, 1\%, 5\%, and 10\% levels. 
\end{adjustbox}
\end{table}

\begin{table}
\centering
\scriptsize
\begin{adjustbox}{minipage=17.5cm, pagecenter}
\vspace{-60pt}
\caption{\textbf{Evaluating Builder Trust}}
\vspace{1em}
        \label{tab:builders}
{
\def\sym#1{\ifmmode^{#1}\else\(^{#1}\)\fi}
\begin{tabular}{@{\extracolsep{2pt}}l*{6}{c}@{}}
\hline\hline

 & \thead{\textsf{Slippage}\\ USDC only (1)} & \thead{\textsf{AdverseSlippage}\\ USDC only (2)} & \thead{\textsf{ReorderingSlippage}\\ USDC only (3)} & \thead{\textsf{Slippage}\\ PEPE only (4)} & \thead{\textsf{AdverseSlippage}\\ PEPE only (5)} & \thead{\textsf{ReorderingSlippage}\\ PEPE only (6)} \\
\hline
\textsf{Public} & -18.5635 & -22.4139 & -37.7070\sym{+} & -6.1565 & -15.5461\sym{*} & -27.1810\sym{**} \\
 & (15.3030) & (15.9791) & (20.3006) & (9.3542) & (7.8429) & (10.4917) \\
\textsf{Flashbots} & 15.0277 & 10.7688 & 19.4199 & 5.4474 & -1.4634 & 1.0724 \\
 & (16.4966) & (17.2254) & (21.8839) & (9.7062) & (8.1380) & (10.8865) \\
\textsf{beaverbuild.org} & 7.3803 & 4.1018 & 7.8514 & 8.2964 & 10.5983 & 25.4452\sym{*} \\
 & (16.6729) & (17.4096) & (22.1179) & (10.0328) & (8.4119) & (11.2529) \\
\textsf{builder0x69} & 6.5001 & 1.9146 & 2.9617 & 1.2382 & 11.8918 & 24.1788\sym{*} \\
 & (17.0873) & (17.8423) & (22.6677) & (10.1194) & (8.4845) & (11.3501) \\
\textsf{rsync-builder.xyz} & 3.4111 & -0.9587 & -1.1272 & 6.0898 & 4.7847 & 17.1691 \\
 & (17.2274) & (17.9886) & (22.8534) & (10.2704) & (8.6111) & (11.5194) \\
\textsf{Titan Builder} & 0.4353 & -2.2372 & -3.7383 & -0.7663 & -0.8555 & -0.6282 \\
 & (22.1798) & (23.1598) & (29.4232) & (13.8219) & (11.5889) & (15.5028) \\
\textsf{Public}:\textsf{Flashbots} & -18.9569 & -14.5283 & -25.3347 & -6.4908 & -1.0603 & -2.7282 \\
 & (16.5873) & (17.3202) & (22.0043) & (9.7748) & (8.1955) & (10.9635) \\
\textsf{Public}:\textsf{beaverbuild.org} & 8.7047 & 11.7424 & 20.8716 & -10.5897 & -13.7707 & -24.6458\sym{*} \\
 & (16.7521) & (17.4922) & (22.2229) & (10.0834) & (8.4543) & (11.3097) \\
\textsf{Public}:\textsf{builder0x69} & -19.3083 & -12.9409 & -18.5683 & -5.7843 & -18.1672\sym{*} & -28.5244\sym{*} \\
 & (17.1636) & (17.9219) & (22.7688) & (10.1669) & (8.5244) & (11.4033) \\
\textsf{Public}:\textsf{rsync-builder.xyz} & -7.5911 & -1.8739 & -5.5371 & -4.8459 & -4.7420 & -18.6074 \\
 & (17.3371) & (18.1031) & (22.9990) & (10.3242) & (8.6562) & (11.5797) \\
\textsf{Public}:\textsf{Titan Builder} & -9.4851 & -6.3563 & -11.3852 & 0.3029 & 0.1535 & -2.4457 \\
 & (22.5443) & (23.5404) & (29.9068) & (13.9849) & (11.7255) & (15.6856) \\
\textsf{orderSize} & -2222.1129\sym{***} & -2107.8426\sym{***} & -3676.2999\sym{***} & -1835.8060\sym{***} & -3179.3877\sym{***} & -6503.4761\sym{***} \\
 & (6.0287) & (6.2951) & (7.9975) & (27.0581) & (22.6866) & (30.3487) \\
\textsf{gasPrice} & 0.0148 & 0.0540\sym{***} & 0.1377\sym{***} & -0.0284\sym{***} & 0.0354\sym{***} & 0.0573\sym{***} \\
 & (0.0129) & (0.0135) & (0.0172) & (0.0047) & (0.0039) & (0.0053) \\
\textsf{logLatency} & -0.9401\sym{+} & -0.4698 & -1.5135\sym{*} & -0.3921 & -0.2385 & -0.2213 \\
 & (0.5130) & (0.5357) & (0.6805) & (0.2785) & (0.2335) & (0.3124) \\
\textsf{slippageTolerance} & -0.0137\sym{***} & -0.0140\sym{***} & -0.0238\sym{***} & -0.0051\sym{***} & -0.0063\sym{***} & -0.0113\sym{***} \\
 & (0.0009) & (0.0009) & (0.0011) & (0.0005) & (0.0004) & (0.0006) \\
\textsf{lastHourReturn} & 0.0196\sym{**} & 0.0714\sym{***} & 0.0904\sym{***} & -0.0036\sym{***} & 0.0000 & -0.0000 \\
 & (0.0071) & (0.0074) & (0.0094) & (0.0005) & (0.0004) & (0.0006) \\
\textsf{liquidity} & -0.1217 & 0.1090 & -0.3161\sym{*} & 4.9746\sym{+} & 4.7885\sym{+} & 8.8098\sym{**} \\
 & (0.1211) & (0.1265) & (0.1607) & (2.9144) & (2.4436) & (3.2688) \\
\textsf{volatility} & 0.0951\sym{***} & 0.1199\sym{***} & 0.2654\sym{***} & -0.0081\sym{***} & -0.0025\sym{*} & -0.0038\sym{**} \\
 & (0.0265) & (0.0277) & (0.0351) & (0.0012) & (0.0010) & (0.0013) \\
Intercept & 44.2156\sym{*} & 39.5196\sym{*} & 76.3465\sym{***} & 14.1824 & 20.6180\sym{**} & 34.1672\sym{**} \\
 & (17.1715) & (17.9302) & (22.7793) & (9.5459) & (8.0036) & (10.7068) \\

\hline
Obs & 258370 & 258370 & 258370 & 169903 & 169903 & 169903 \\
Adj. R\sym{2} & 0.3487 & 0.3059 & 0.4527 & 0.0292 & 0.1046 & 0.2139 \\
F-stat & 2944.2481 & 2423.2304 & 4548.6162 & 151.5305 & 585.0631 & 1360.5539 \\
\hline\hline
\end{tabular}
}
\footnotesize

In this table, we present coefficients for regressions for the model in Equation~\ref{eq:mainregression}, but where the endogenous variables are dollar notion of slippage, namely $y_i := \mathsf{Slippage}_i \cdot \mathsf{orderSize}_i$. We also add two new variables. First, $\mathsf{Public}_i$ is a dummy variable that denotes whether swap $i$ is first seen in the public mempool. Second, $\mathsf{Builder}_i$ is set to the name of the builder that built the block containing swap $i$, if the builder is one of the top 5 builders by total number of private transactions in the sample, else it is set to `Other'. 
We further interact $\mathsf{Public}$ with $\mathsf{Builder}$. Note that most swaps do not experience adversarial slippage, so here we use a dollar notion of slippage may help to amplify the effect of adversarial slippage.
Models (1) and (2) are run for the USDC-WETH dataset, whereas models (3) and (4) are run for the PEPE-WETH dataset. All regressions include weekly fixed effects. ***, **, *, and $^+$ denote statistical significance at the 0.1\%, 1\%, 5\%, and 10\% levels.
\end{adjustbox}
\end{table}

\clearpage
\section{Additional Charts}
\label{sec:breakdowncost}

In this section, we present some additional charts that we did not have a chance to present in the main body. They characterize overall confirmation times, as well as transaction cost breakdowns, and may be useful as a reference.
\begin{table}[]
    \centering
     \caption{\textbf{Transaction Confirmation Time Percentiles}}
     \vspace{1em}
         \label{tab:my_label}
\begin{tabular}{ll}
\toprule
Percentile &      Seconds \\
\midrule
50th   &    6.2 \\
80th   &   10.2 \\
90th   &   12.1 \\
95th   &   17.9 \\
97th  &   24.1 \\
99th  &  104.0 \\
99.5th &  259.9 \\
\bottomrule
\end{tabular}
    \footnotesize
    
     \vspace{1em}
The transaction confirmation time is defined as the difference between user signature time and the timestamp of the block in which the transaction is realized. For context, the time interval between two consecutive blocks in Ethereum (assume no missing block) is 12 seconds. So roughly 90\% of all transactions get confirmed immediately in the next block after the users sign and broadcast their transaction.
\end{table}

\begin{figure}[h!]
    \centering
    \caption{\textbf{Transaction Cost Composition For Different Transaction Types}}
    \begin{subfigure}[b]{0.45\textwidth}
        \includegraphics[width=\textwidth]{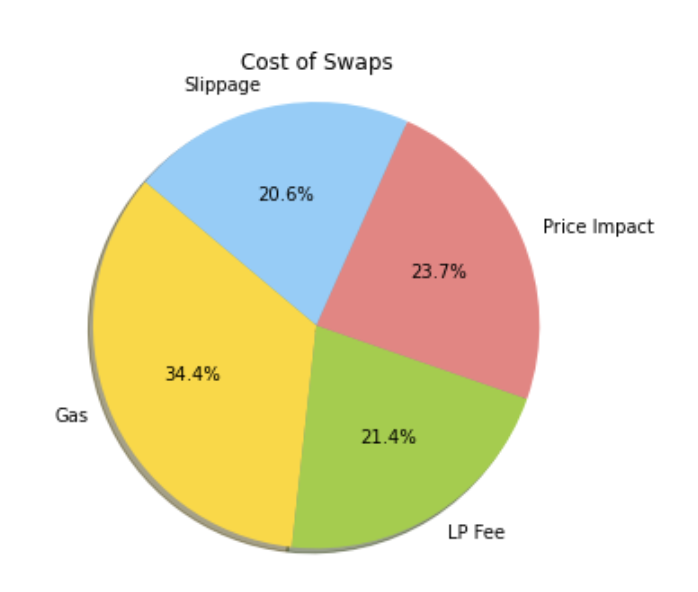}
        \caption{Full Sample}
        \label{fig:pie_charts:fullsample}
    \end{subfigure}
    \hfill
    \begin{subfigure}[b]{0.45\textwidth}
        \includegraphics[width=\textwidth]{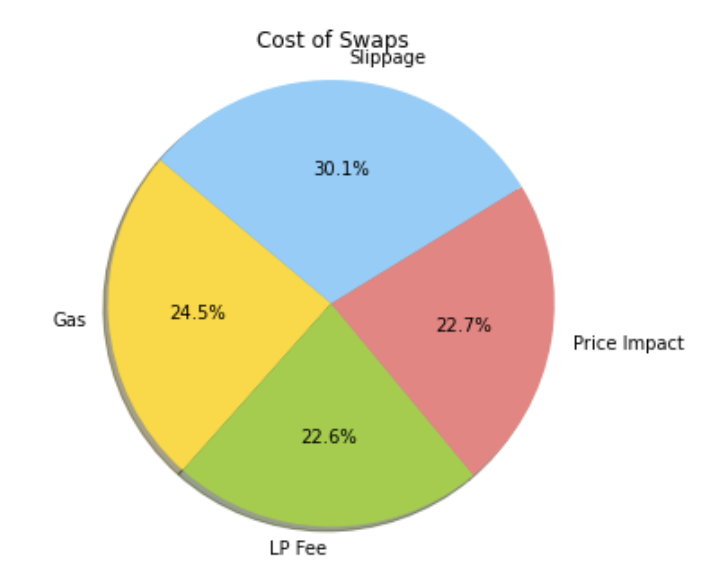}
        \caption{WETH-USDC Swaps}
    \end{subfigure}
    \vspace{1em} 
    \begin{subfigure}[b]{0.45\textwidth}
        \includegraphics[width=\textwidth]{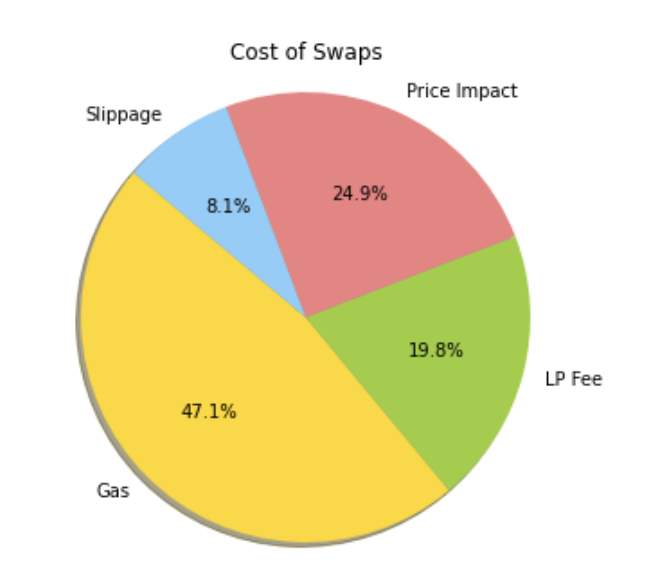}
        \caption{WETH-PEPE Swaps}
    \end{subfigure}
    \hfill
    \begin{subfigure}[b]{0.45\textwidth}
        \includegraphics[width=\textwidth]{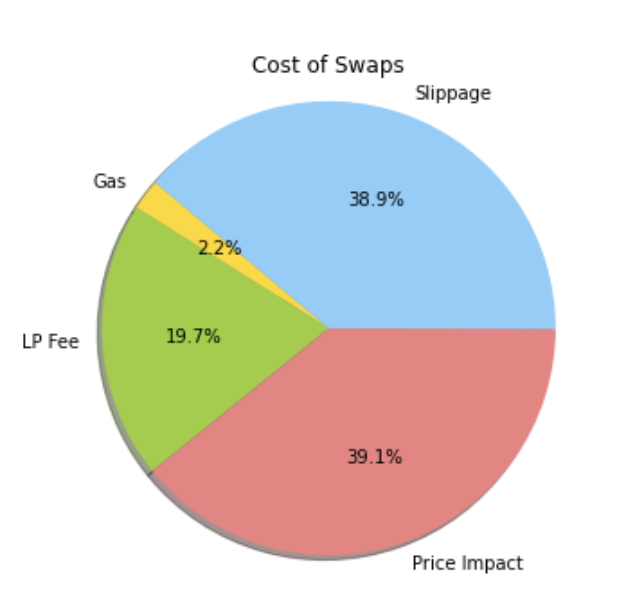}
        \caption{Large-sized Swaps ($\geq$\$100k)}
    \end{subfigure}
    \hfill
    \begin{subfigure}[b]{0.45\textwidth}
        \includegraphics[width=\textwidth]{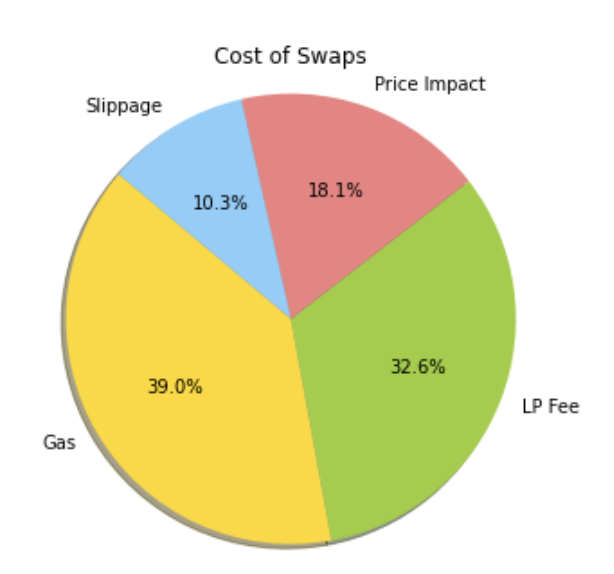}
        \caption{Medium-sized Swaps (\$1k - \$100k)}
    \end{subfigure}
    \hfill
    \begin{subfigure}[b]{0.45\textwidth}
        \includegraphics[width=\textwidth]{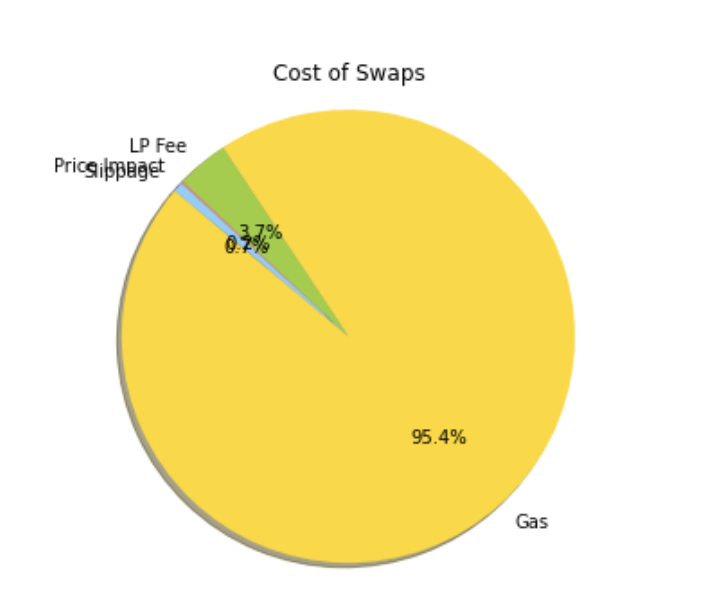}
        \caption{Small-sized Swaps ($\leq$\$1k)}
    \end{subfigure}
    \label{fig:pie_charts}
    \footnotesize
    \vspace{1em}
    
    Here we show the percentage composition of the transaction costs within different groups. 
\end{figure}
\fi


\ignore{
\pagebreak
\section{Unused regressions}

    \begin{table}[]
    \tiny
        \centering
        \begin{adjustbox}{max width = \dimexpr\paperwidth-3cm, center}
        \input{regressions/ALL_interactive_smalllarge}
        \end{adjustbox}
        \caption{PublicVar table, with interactions, all swaps, USDC and PEPE, bps. Small (25 percentile), Large (75 percentile). Week fixed effects hidden}
    \end{table}
    
    \begin{table}[]
    \tiny
        \centering
        \begin{adjustbox}{max width = \dimexpr\paperwidth-3cm, center}
        \input{regressions/ALL_noninteractive_smalllarge}
        \end{adjustbox}
        \caption{PublicVar table, without interactions, all swaps, USDC and PEPE, bps. Small (25 percentile), Large (75 percentile). FE week hidden.}
    \end{table}

}
        
\end{document}